\documentclass{article}  
\usepackage{graphicx}
\usepackage{txfonts}
\begin{document}
\noindent
{\em 2021 outburst of RS Oph  \hfill{29 August 2021}}
\begin{center}
\vskip 0.5 cm
{\large \bf The 2021 outburst of RS Oph. A pictorial atlas of the spectroscopic evolution: the first 18 days}\\
\vskip 0.7 cm
{\large Ulisse Munari$^1$ and Paolo Valisa$^2$}
\end{center}
\vskip 0.3 cm
\noindent
\phantom{~~~~~~~~~~~}{\small 1: INAF National Institute of Astrophysics, 36012 Asiago, Italy}\\
\phantom{~~~~~~~~~~~}{\small 2: ANS Collaboration, c/o Astronomical Observatory, 36012 Asiago, Italy}\\
\vskip 0.5 cm

\baselineskip 9pt
\begin{center}
\parbox{13.4cm}{\baselineskip 9pt
{\bf Abstract}. {\small \em
A pictorial atlas of the spectroscopic evolution at optical wavelengths
is presented for the first 18 days of the 2021 outburst of the recurrent
nova RS Oph,  prior to the emergence of high ionization emission lines.
The spectra presented here have been obtained at daily cadence with the
Asiago 1.22m + B\&C (3200-7900 Ang, 2.3 \AA/pix) and Varese 0.84m
+ Echelle telescopes (4250-8900 \AA, resolving power 18,000). The spectra
have been fully calibrated in IRAF, absolutely fluxed, and heliocentric
corrected. The Echelle spectra have been also corrected for telluric absorptions.}}
\end{center}
\vskip 0.3 cm
\baselineskip 13pt

\noindent
\underline{\sc INTRODUCTION}
\bigskip

The recurrent nova and symbiotic binary RS Oph underwent a new outburst in
August 2021, following previous ones recorded in 1898, 1933, 1958, 1967,
1985 and 2006.  At each new event a better coverage of the eruption was
obtained with observations progressively expanding to cover the whole
electromagnetic spectrum.  For the 2006 event, an impressive amount of data
has been gathered, and the book {\em RS Ophiuchi (2006) and the Recurrent
Nova Phenomenon} edited by Bode, O'Brien \& Darnley (2008, ASPC 401)
provides a detailed summary of the main results.

Yet, a comprehensive and detailed mapping of the spectral evolution at
optical wavelengths seems missing, at leat at the level of completeness,
detail and homogeneity of the present atlas.  It aims to fill the gap, by
providing a quick pictorial guide to the spectral evolution of RS Oph during
the first 18 days of the 2021 outburst.  To make the atlas promptly
available, we just describe how the observations have been collected and how
the reduced spectra have been incorporated in the atlas.  This is of course
no substitute for a proper analysis, to be carried out in due time.  The
latter will benefit from continued observations covering the rest of the
evolution of RS Oph as well as comparison with the results of observations
carried out at other wavelengths (and not yet published).

While presenting this first batch of spectra, our monitoring of RS Oph will
continue at daily cadence for as long as possible, and updates to this atlas
are planned for later epochs.

The current atlas covers the first 18 days of the eruption (August 8 to 26),
those characterized by low ionization conditions.  From day +19 (August 27),
the evolution has progressed to the emergence of higher ionization lines
(HeII, [FeVII]) as reported by Shore et al.  (2021b,c). 

The current outburst has triggered observations of RS Oph over the whole
electromagnetic spectrum.  Preliminar reports have been presented by Cheung
et al.  (2021a,b) and Wagner \& HESS Collaboration (2021a,b) for
$\gamma$-rays; Enoto et al.  (2021a,b), Ferrigno et al.  (2021), Luna et al. 
(2021), Page (2021), Page et al.  (2021), Rout et al.  (2021) and Shidatsu
et al.  (2021) for the X-rays; Mikolajewska et al.  (2021), Munari U.,
Valisa P., (2021a,b), Shore et al.  (2021a,b,c), and Taguchi et al. 
(2021a,b) for the optical; Woodward et al.  (2021) for the infrared; and
Sokolovsky et al.  (2021) and Williams et al.  (2021) for the radio.  The
results of the search for neutrino emission has been given by Pizzuto et al. 
(2021), and on the polarization of optical light by Nikolov \& Luna (2021). 
A great summary of the main results obtained on the previous outburst of RS
Oph in 2006 is the conference proceedings edited by Evans et al. (2008).
\bigskip

\noindent
\underline{\sc OBSERVATIONS}
\bigskip

RS Oph has been observed each day for the first 18 days of its 2021 eruption
with the Echelle slit-spectrograph mounted on the Varese 0.84m telescope. 
The spectra have been recorded over the range 4250-8900 \AA, at a 18,000
resolving power.  The data reduction has been performed in IRAF and has
included all usual steps of correction for bias, dark and flat, sky
subtraction, wavelength calibration (via Thorium lamp), and heliocentric
correction.  Spectrophotometric standards, located close on the sky to RS
Oph, have been observed each night soon before and after the nova to achieve
an optimal flux calibration, which was also essential to accurately merge
the individual 30 Echelle orders into a single 1D fluxed spectrum not
affected by dents at the points of jointure.  Telluric dividers were
observed each night similarly to the spectrophotometric standards, to
properly remove from the spectra of RS Oph the presence of telluric
absorption lines.  A journal of the Echelle observations is provided in
Table~1.  The poor quality of the spectrum for Aug 12 is due to cloud cover
persisting all night.

RS Oph was also observed on almost each day for the first 18 days of its
2021 eruption with the B\&C spectrograph attached to the Asiago 1.22m
telescope, allowing to cover the 3200-7900 \AA\ range at 2.3 \AA/pix
dispersion.  The B\&C spectra have been equally and fully reduced in IRAF
and flux calibrated against spectrophotometric standards observed each night
at a similar height on the horizon as RS Oph.  A journal of the B\&C
observations is presented in Table~2.

At both telescopes a great emphasis has been put on maintaining the highest
homogeneity throughout the observing campaign on the instrument set-up,
choice of standards, observation and data reduction procedures.  Any
difference between the spectra presented in this atlas is more intrinsic to
RS Oph than due to any stage of the observing process.

\bigskip

\noindent
\underline{\sc THE ATLAS}
\bigskip

Given the huge intensity of the emission lines displayed by RS Oph and the
necessity to compact many different spectra on the same picture, we adopt to
plot the spectra on all figure as the {\bf logarithm of the flux} (in erg
cm$^{-2}$ s$^{-1}$ \AA$^{-1}$).  Multi-epoch observations within the same
night have been averaged into one single spectrum per night.

The sequence of B\&C spectra is presented first, with Figure~1 covering the
whole wavelength interval, and Figures~2 and 3 zooming on the blue and red
portions respectively, with identification of the principal lines.  A list
of the emission lines identified in the spectrum for Aug 23
($t-t_\circ$=+15.28 days) is given in Table~3.

Selected portions of the Echelle spectra then follow. In Figure~4 we present
the evolution of H$\alpha$, and that of HeI 5876 follows in Figure~5. Figures~6 and
7 are devoted to cover H$\beta$ and the nearby FeII multiplet 42 and HeI
lines at 4921 and 5015 \AA. Finally, Figure~8, zooms on the sharp components displayed by H$\alpha$,
H$\beta$, FeII \#42 5018 and HeI 5015 on top of the much wider and stronger 
emission lines, and to the NaI 5890, 5896 \AA\ doublet.
\bigskip

\noindent
\underline{\sc REFERENCE EPOCH}
\bigskip

To identify the spectra through the various figures, we have labelled them
according to the observing UT date as listed in Tables~1 and 2. In the same 
tables the time elapsed since the onset of the outburst ($t- t_\circ$) 
and since maximum brightness ($t-t_{\rm max}$) in the $V$-band are listed.

The very quick rise to maximum has been covered by L.P. Lou and B. Wang
with observations obtained with a Sony DSLR camera and reported to AAVSO
(observer code WPIA). Extrapolating back their linear rise to intercept
the $V$=11.1 mag quiescence brightness that RS Oph displayed on 
preceding days, set the time of eruption to:
\begin{equation}
t_\circ = 2459435.00~{\rm JD}  ~~~~~~ 2021 ~{\rm Aug}~ 08.50   ~~~~(\pm 0.01)
\end{equation}
The time of passage at maximum $V$-band magnitude is less accurately determined.
Again from the AAVSO light-curve, we estimate:
\begin{equation}
t^{V}_{\rm max} = 2459436.18~{\rm JD}  ~~~~~~ 2021 ~{\rm Aug}~ 09.58   ~~~~(\pm 0.05)
\end{equation}
\bigskip

\noindent
\underline{\sc ACKNOWLEDGMENTS}
\bigskip

We express our gratitude to P. Ochner (Univ. Padova) and A. Vagnozzi (ANS
Collaboration) for their help with the acquisition of some of the spectra
used in this paper.
\bigskip

\noindent
\underline{\sc REFERENCES}
\bigskip

Cheung C.~C., Ciprini S., Johnson T.~J., 2021a, ATel, 14834

Cheung C.~C., Johnson T.~J., Mereu I., et al., 2021b, ATel, 14845

Enoto T., Maehara H., Orio M., et al., 2021a, ATel, 14850

Enoto T., Orio M., Fabian A., et al., 2021b, ATel, 14864

Evans A., Bode M. F., O'Brien T. J., Darnley M. J., eds., 2008, RS Ophiuchi (2006) and the Recurrent Nova 

{~~~~~}Phenomenon, ASPC, 401  

Ferrigno C., Savchenko V., Bozzo E., et al., 2021, ATel, 14855

Luna G.~J.~M., Jimemez-Carrera R., Enoto T., et al., 2021, ATel, 14872

Mikolajewska J., Aydi E., Buckley D., et al., 2021, ATel, 14852

Munari U., Valisa P., 2021a, ATel, 14840

Munari U., Valisa P., 2021b, ATel, 14860

Nikolov Y., Luna G.~J.~M., 2021, ATel, 14863

Page, K.~L. 2021, ATel, 14885

Page K.~L., Osborne J.~P., Aydi E., 2021, ATel, 14848

Pizzuto A., Vandenbroucke J., Santander M., IceCube Collaboration, 2021, ATel, 14851

Rout S.~K., Srivastava M.~K., Banerjee D.~P.~K., et al., 2021, ATel, 14882

Shidatsu M., Negoro H., Mihara T., et al., 2021, ATel, 14846

Shore S.~N., Allen H., Bajer M., et al., 2021a, ATel, 14868

Shore S.~N., Teyssier F., Thizy O.,  2021b, ATel, 14881

Shore S.~N., Teyssier F., Guarro J., et al., 2021c, ATel, 14883

Sokolovsky K., Aydi E., Chomiuk L., et al., 2021, ATel, 14886

Taguchi K., Ueta, T., Isogai, K., 2021a, ATel, 14838

Taguchi K., Maheara H., Isogai K., et al., 2021b, ATel, 14858

Williams D., O'Brien T., Woudt P., et al., 2021, ATel, 14849

Wagner, S.~J., HESS Collaboration, 2021a, ATel, 14844

Wagner, S.~J., HESS Collaboration, 2021b, ATel, 14857

Woodward C.~E., Evans A., Banerjee D.~P.~K., et al., 2021, ATel, 14866

\bigskip

\begin{center}
\begin{table}[!b]
\parbox{9.4cm}{
\caption{Journal of Echelle spectroscopic observations obtained 
with the Varese 0.84m telescope.}
\begin{tabular}{ccrcrr}
&&\\
\hline
&&\\
\multicolumn{2}{c}{date UT} & expt  & HJD        &  $t-t_\circ$ & $t-t_{\rm max}$ \\ \cline{1-2}
yyyy-mm-dd & hh:mm    & (sec) & (-2459400) & (days)       & (days)          \\
&&\\
2021-08-09 & 19:30 &  900  &  36.316  &    1.32   &  0.14  \\
2021-08-09 & 20:03 &  540  &  36.339  &    1.34   &  0.16  \\
2021-08-09 & 20:55 &  540  &  36.375  &    1.38   &  0.20  \\
2021-08-09 & 21:41 &  360  &  36.407  &    1.41   &  0.23  \\
2021-08-10 & 19:41 &  720  &  37.324  &    2.32   &  1.14  \\
2021-08-10 & 21:21 &  720  &  37.393  &    2.39   &  1.21  \\
2021-08-11 & 19:23 &  720  &  38.311  &    3.31   &  2.13  \\
2021-08-11 & 20:11 & 1200  &  38.344  &    3.34   &  2.16  \\
2021-08-12 & 19:50 & 1200  &  39.330  &    4.33   &  3.15  \\
2021-08-13 & 19:31 &  900  &  40.316  &    5.32   &  4.14  \\
2021-08-14 & 20:40 & 1440  &  41.364  &    6.36   &  5.18  \\
2021-08-15 & 19:40 & 1200  &  42.323  &    7.32   &  6.14  \\
2021-08-16 & 20.05 & 1500  &  43.338  &    8.34   &  7.16  \\
2021-08-17 & 19:20 &  900  &  44.309  &    9.31   &  8.13  \\
2021-08-18 & 19:30 & 1500  &  45.315  &   10.32   &  9.14  \\
2021-08-19 & 20:31 & 1500  &  46.358  &   11.36   & 10.18  \\
2021-08-20 & 19:28 & 1500  &  47.314  &   12.31   & 11.13  \\
2021-08-21 & 19:18 & 1500  &  48.307  &   13.31   & 12.13  \\
2021-08-22 & 19:20 & 1500  &  49.308  &   14.31   & 13.13  \\
2021-08-23 & 19:45 & 2400  &  50.325  &   15.33   & 14.15  \\
2021-08-24 & 19:41 & 1800  &  51.323  &   16.32   & 15.14  \\
2021-08-25 & 19:15 & 1800  &  52.304  &   17.30   & 16.12  \\
2021-08-26 & 19:10 & 2400  &  53.301  &   18.30   & 17.12  \\
&&\\
\hline
\end{tabular}}
\end{table}
\end{center}

\begin{center}
\begin{table}
\parbox{9.4cm}{
\caption{Journal of B\&C spectroscopic observations obtained 
with the Asiago 1.22m telescope.}
\begin{tabular}{ccrcrr}
&&\\
\hline
&&\\
\multicolumn{2}{c}{date UT} & expt  & HJD        &  $t-t_\circ$ & $t-t_{\rm max}$ \\ \cline{1-2}
yyyy-mm-dd & hh:mm    & (sec) & (-2459400) & (days)       & (days)          \\
&&\\
 2021-08-09 & 18:49 & 100 & 36.284  &   1.28   &  0.10  \\
 2021-08-09 & 19:56 & 150 & 36.331  &   1.33   &  0.15  \\
 2021-08-10 & 19:11 & 120 & 37.300  &   2.30   &  1.12  \\
 2021-08-10 & 21:11 & 150 & 37.383  &   2.38   &  1.20  \\
 2021-08-11 & 19:09 &  90 & 38.298  &   3.30   &  2.12  \\
 2021-08-11 & 22:03 & 130 & 38.419  &   3.42   &  2.24  \\
 2021-08-12 & 18:44 &  90 & 39.281  &   4.28   &  3.10  \\
 2021-08-12 & 19:05 & 170 & 39.295  &   4.30   &  3.12  \\
 2021-08-12 & 20:45 & 120 & 39.365  &   4.37   &  3.19  \\
 2021-08-13 & 19:39 & 120 & 40.319  &   5.32   &  4.14  \\
 2021-08-13 & 21:04 & 120 & 40.378  &   5.38   &  4.20  \\
 2021-08-14 & 20:24 & 100 & 41.351  &   6.35   &  5.17  \\
 2021-08-15 & 20:11 & 120 & 42.342  &   7.34   &  6.16  \\
 2021-08-15 & 21:16 &  90 & 42.386  &   7.39   &  6.21  \\
 2021-08-17 & 19:13 & 120 & 44.301  &   9.30   &  8.12  \\
 2021-08-18 & 19:24 & 120 & 45.309  &  10.31   &  9.13  \\
 2021-08-19 & 18:44 & 110 & 46.281  &  11.28   & 10.10  \\
 2021-08-19 & 19:54 &  90 & 46.330  &  11.33   & 10.15  \\
 2021-08-20 & 19:22 & 120 & 47.307  &  12.31   & 11.13  \\
 2021-08-20 & 20:51 &  90 & 47.369  &  12.37   & 11.19  \\
 2021-08-22 & 19:42 & 150 & 49.321  &  14.32   & 13.14  \\
 2021-08-23 & 18:41 & 150 & 50.279  &  15.28   & 14.10  \\
&&\\
\hline
\end{tabular}}
\end{table}
\end{center}

\begin{center}
\begin{table}
\parbox{14.4cm}{
\caption{Identification of emission lines in the Asiago 1.22m + B\&C spectrum of RS
Oph for Aug 23 ($t-t_\circ$=+15.28 days).}
\begin{tabular}{cl|cl|cl|cl}
\multicolumn{8}{c}{}\\
\hline
&&&&&&& \\
$\lambda$ & line & $\lambda$ & line & $\lambda$ & line & $\lambda$ & line  \\
&&&&&&& \\
 3721        &  H14           &    4176        &  FeII 27+28    &   4823        &  HeI, FeII 42  &   5676        &  NII           \\
 3733        &  H13           &    4233        &  FeII 27       &   4949        &  [OIII]        &   5755        &  [NII]         \\
 3750        &  H12           &    4273        &  FeII 27       &   5007        &  [OIII]        &   5876        &  HeI           \\
 3770        &  H11           &    4300        &  FeII 27+28    &   5019        &  HeI+FeII 42   &   5991        &  FeII 46       \\
 3798        &  H10           &    4340        &  H$\gamma$     &   5048        &  HeI           &   6148        &  FeII 74       \\
 3837        &  H9            &    4363        &  [OIII]        &   5169        &  FeII 42       &   6248        &  FeII 74       \\
 3856        &  HeI           &    4388        &  HeI           &   5185        &  FeII          &   6318        &  FeII          \\
 3868        &  HeI           &    4417        &  FeII 27       &   5198        &  FeII 49       &   6369        &  blend         \\
 3889        &  H8+HeI        &    4471        &  HeI           &   5234        &  FeII 49       &   6433        &  FeII 40       \\
 3927        &  HeI           &    4489        &  FeII 37       &   5255        &  FeII 49       &   6456        &  FeII 74       \\
 3936        &  HeI           &    4520        &  FeII 37+38    &   5265        &  FeII 48       &   6563        &  H$\alpha$     \\
 3970        &  H$\epsilon$   &    4555        &  FeII 37+38    &   5276        &  FeII 49       &   6678        &  HeI           \\
 4026        &  HeI           &    4584        &  FeII 38       &   5316        &  FeII 48+49    &   7065        &  HeI           \\
 4070        &  FeII 22       &    4634        &  NIII+FeII 37  &   5363        &  FeII 48       &   7281        &  HeI           \\
 4101        &  H$\delta$     &    4647        &  CIII          &   5426        &  FeII 49       &   7711        &  FeII 73       \\
 4124        &  FeII 22       &    4713        &  HeI           &   5496        &  FeII          &   7774        &  OI            \\
 4144        &  HeI           &    4861        &  H$\beta$      &   5535        &  FeII 55                                        \\
&&&&&&& \\
\hline
\end{tabular}}
\end{table}
\end{center}

\clearpage
\begin{figure}
\includegraphics[width=18cm]{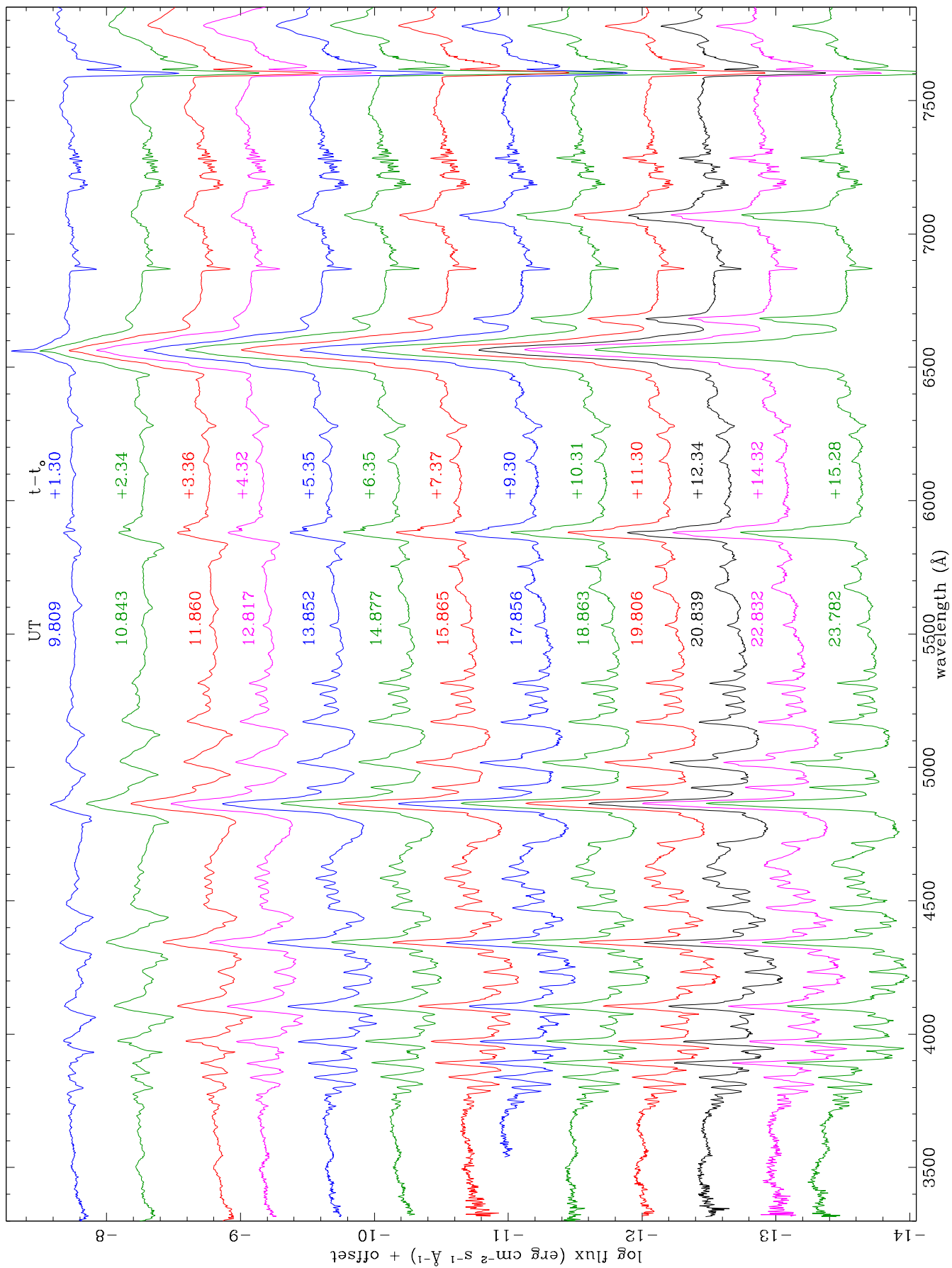}
\caption{Sequence of Asiago 1.22m spectra showing the whole recorded range. 
The logarithm of the flux is adopted as the ordinates and a shift is applied to avoid overplotting.}
\end{figure}

\clearpage
\begin{figure}
\includegraphics[width=18cm]{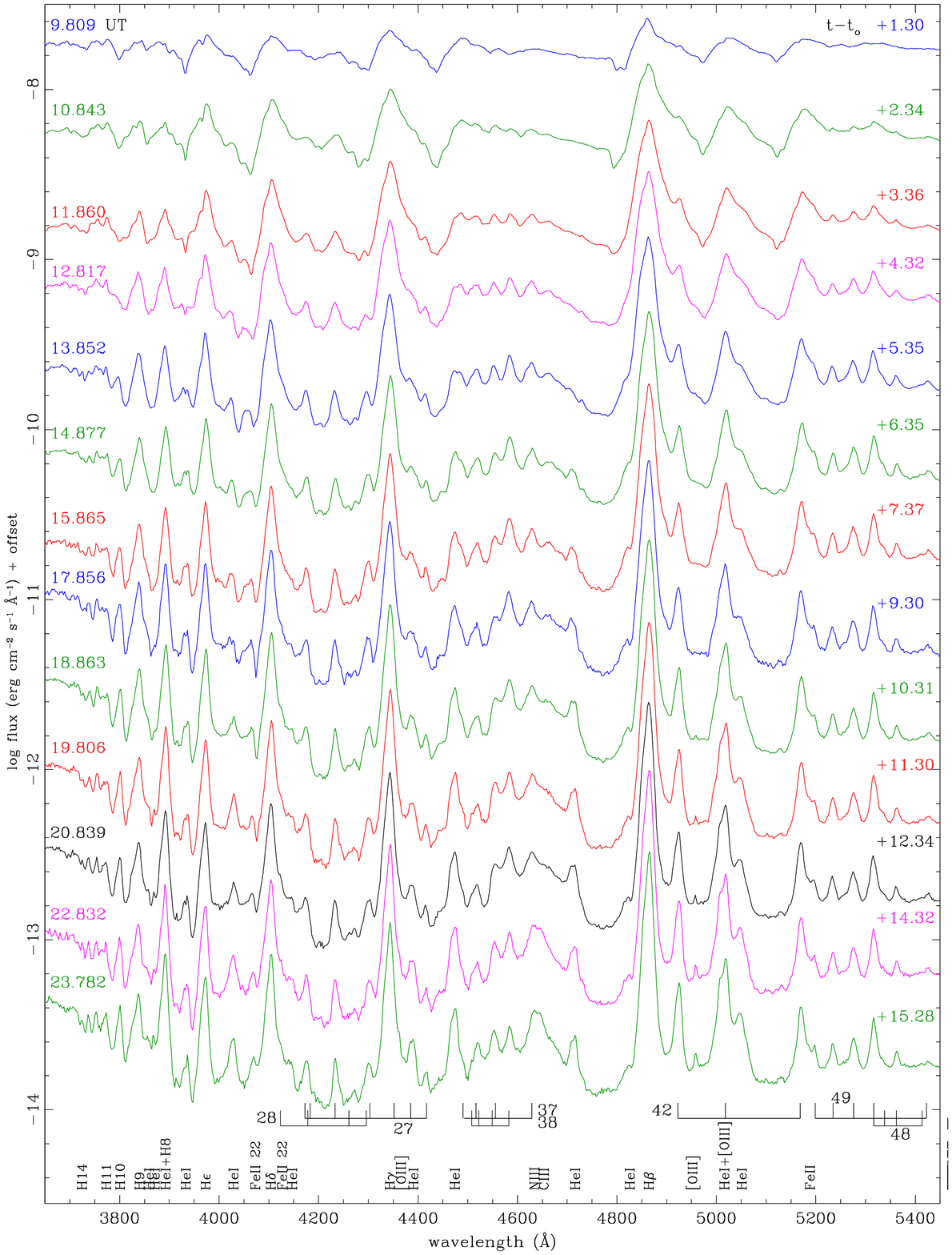}
\caption{An expanded view of the 3650-5450 \AA\ interval for the same spectra of Figure~1.
The logarithm of the flux is adopted as the ordinates and a shift is applied to avoid overplotting.}
\end{figure}

\clearpage
\begin{figure}
\includegraphics[width=18cm]{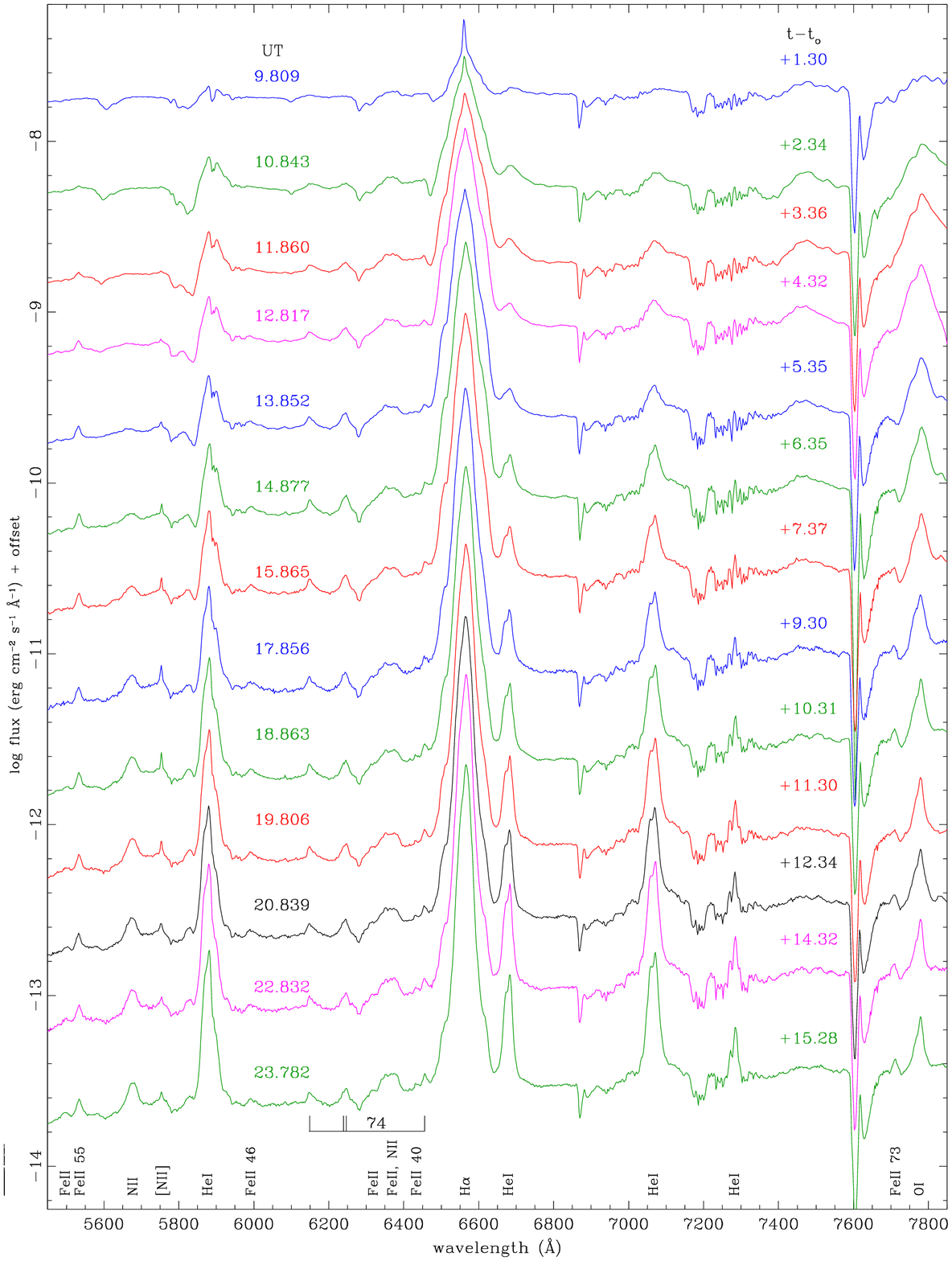}
\caption{An expanded view of the 5450-7850 \AA\ interval for the same spectra of Figure~1.
The logarithm of the flux is adopted as the ordinates and a shift is applied to avoid overplotting.}
\end{figure}

\clearpage
\begin{figure}
\includegraphics[angle=270,width=18cm]{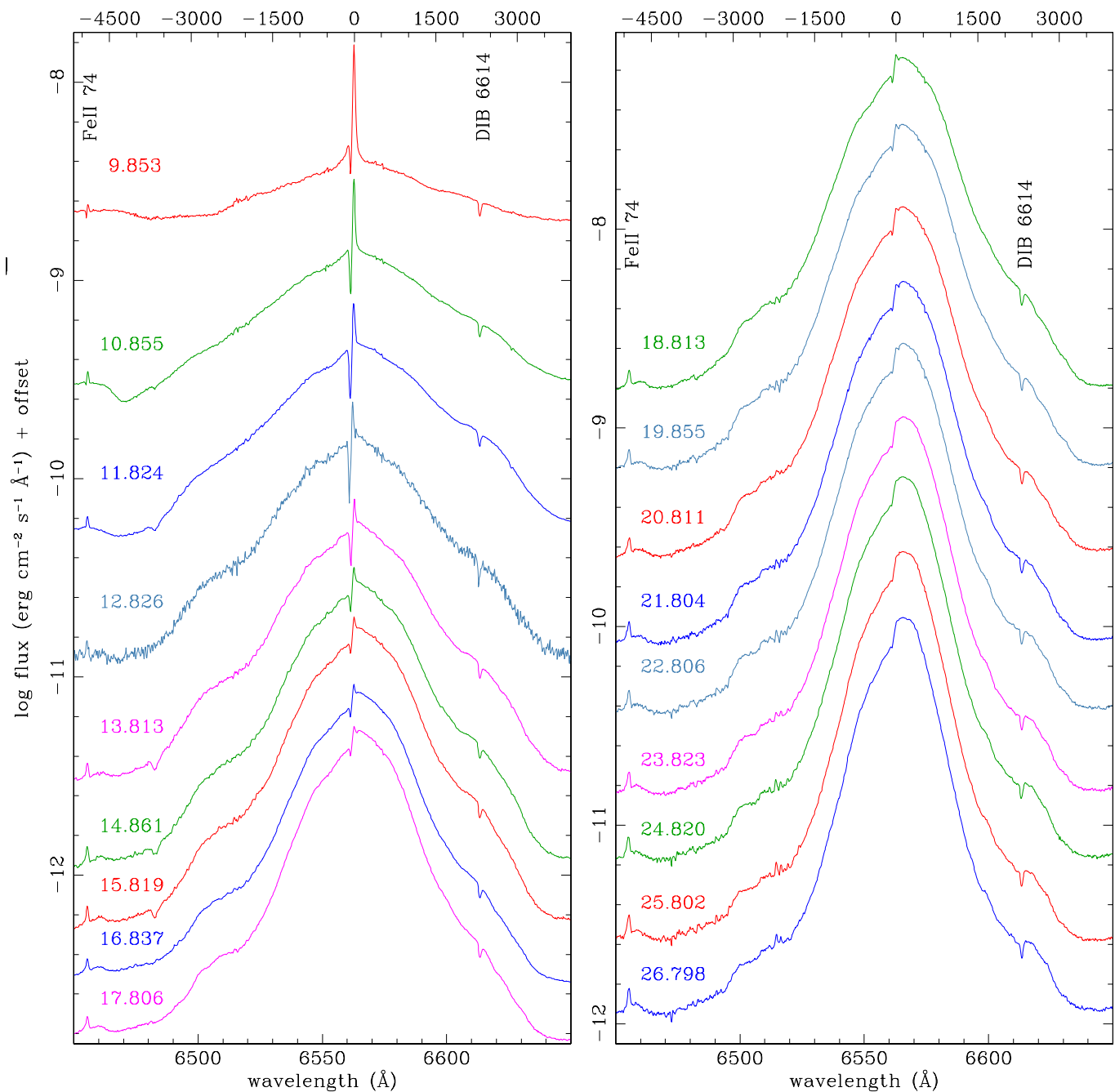}
\caption{Evolution of the H$\alpha$ profile from Varese 0.84m Echelle spectra.
The abscissae at the top give the  heliocentric radial velocity (in
km/s). The logarithm of the flux is adopted as the ordinates and a shift is applied to avoid overplotting.}
\end{figure}

\clearpage
\begin{figure}
\includegraphics[angle=270,width=18cm]{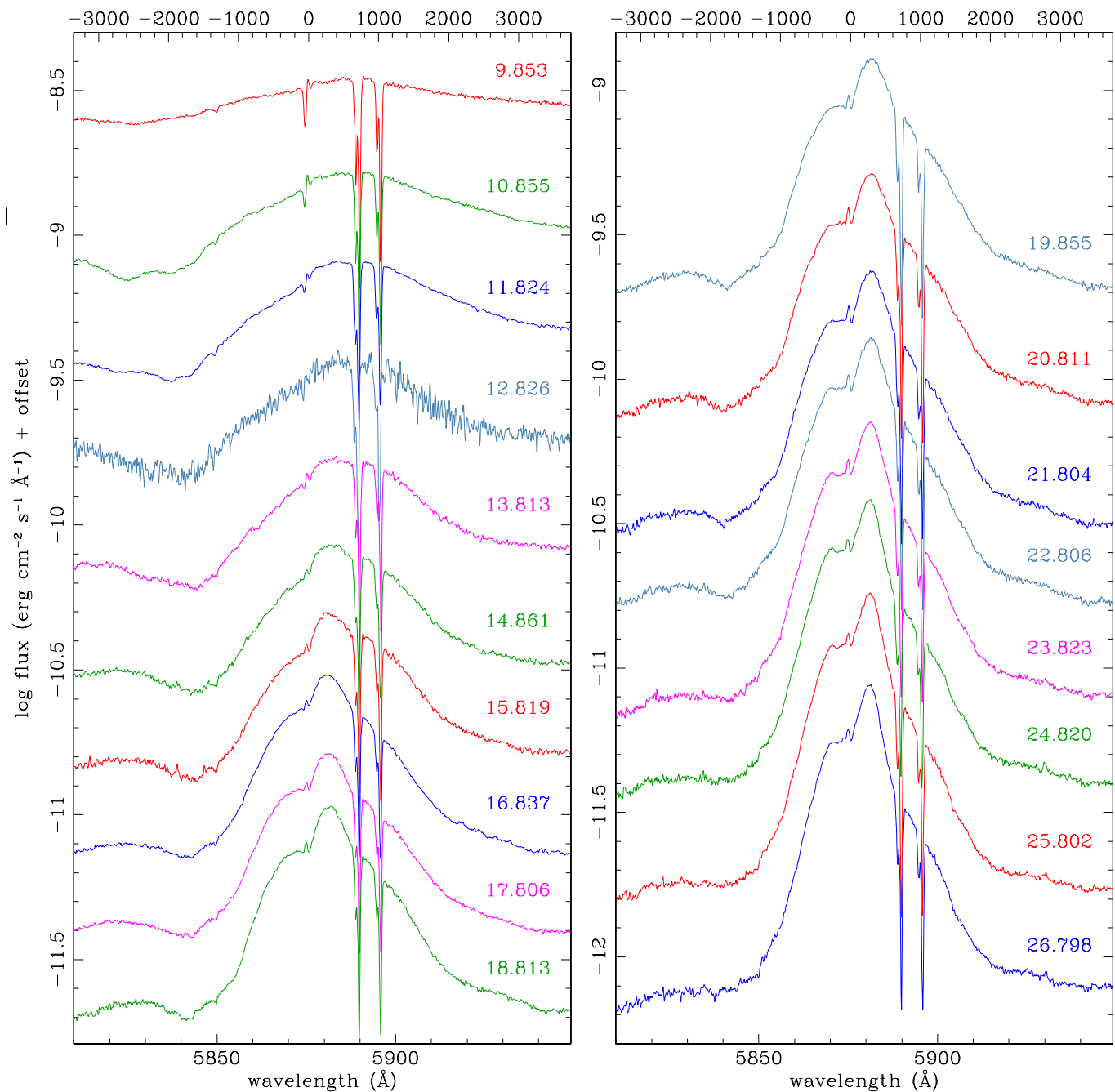}
\caption{Evolution of the HeI 5876 profile and superimposed NaI doublet from Varese 0.84m Echelle spectra.
The abscissae at the top give the  heliocentric radial velocity (in
km/s). The logarithm of the flux is adopted as the ordinates and a shift is applied to avoid overplotting.}
\end{figure}

\clearpage
\begin{figure}
\includegraphics[width=18cm]{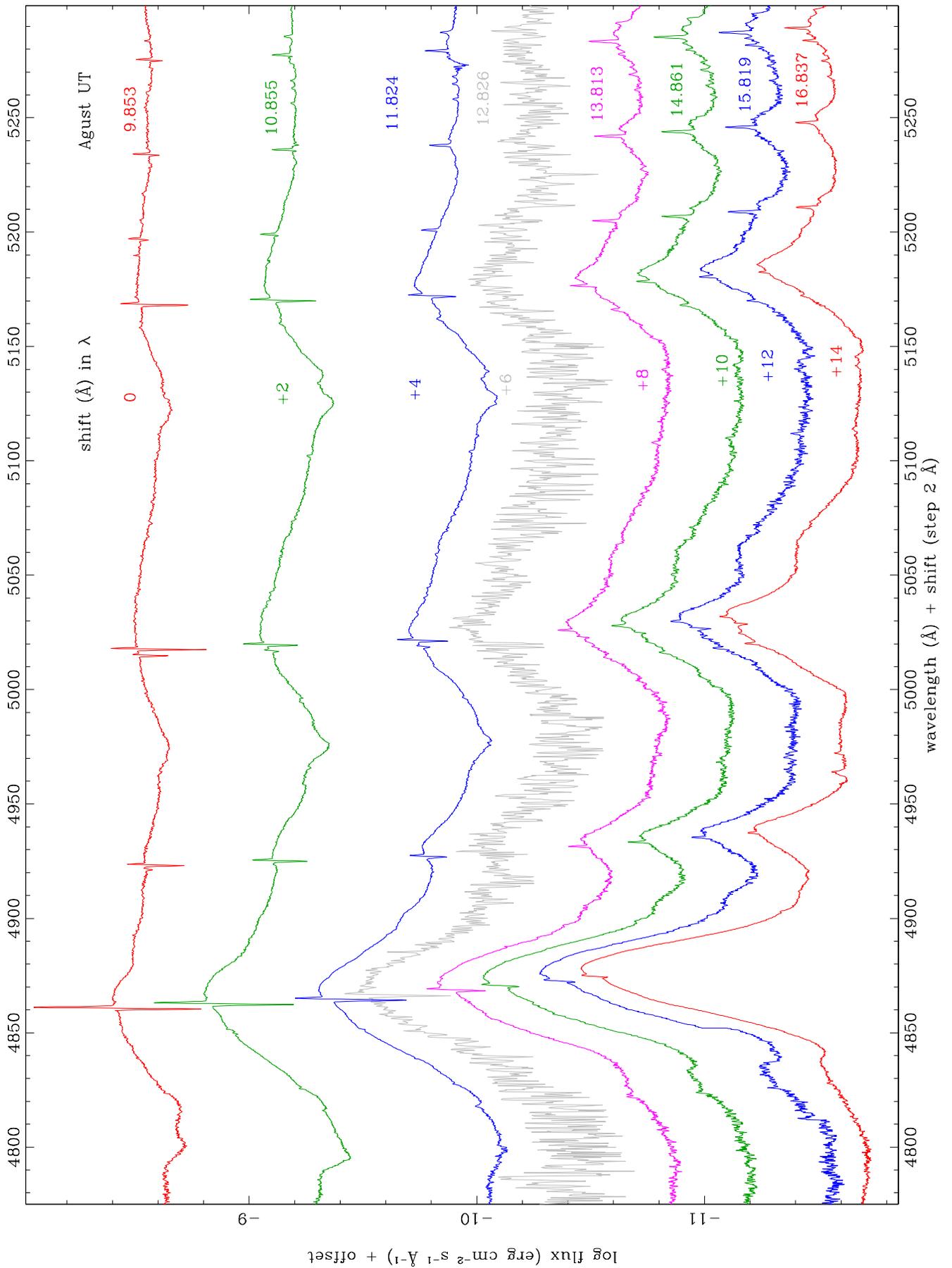}
\caption{Series of Varese 0.84m Echelle spectra covering the H$\beta$, FeII multiplet 42 and HeI 4941, 5015 
emission lines for the first eight days of the monitoring (the remaining dates are covered by the following Figure~7).
The logarithm of the flux is adopted as the ordinates and a shift is applied to avoid overplotting.}
\end{figure}

\clearpage
\begin{figure}
\includegraphics[width=18cm]{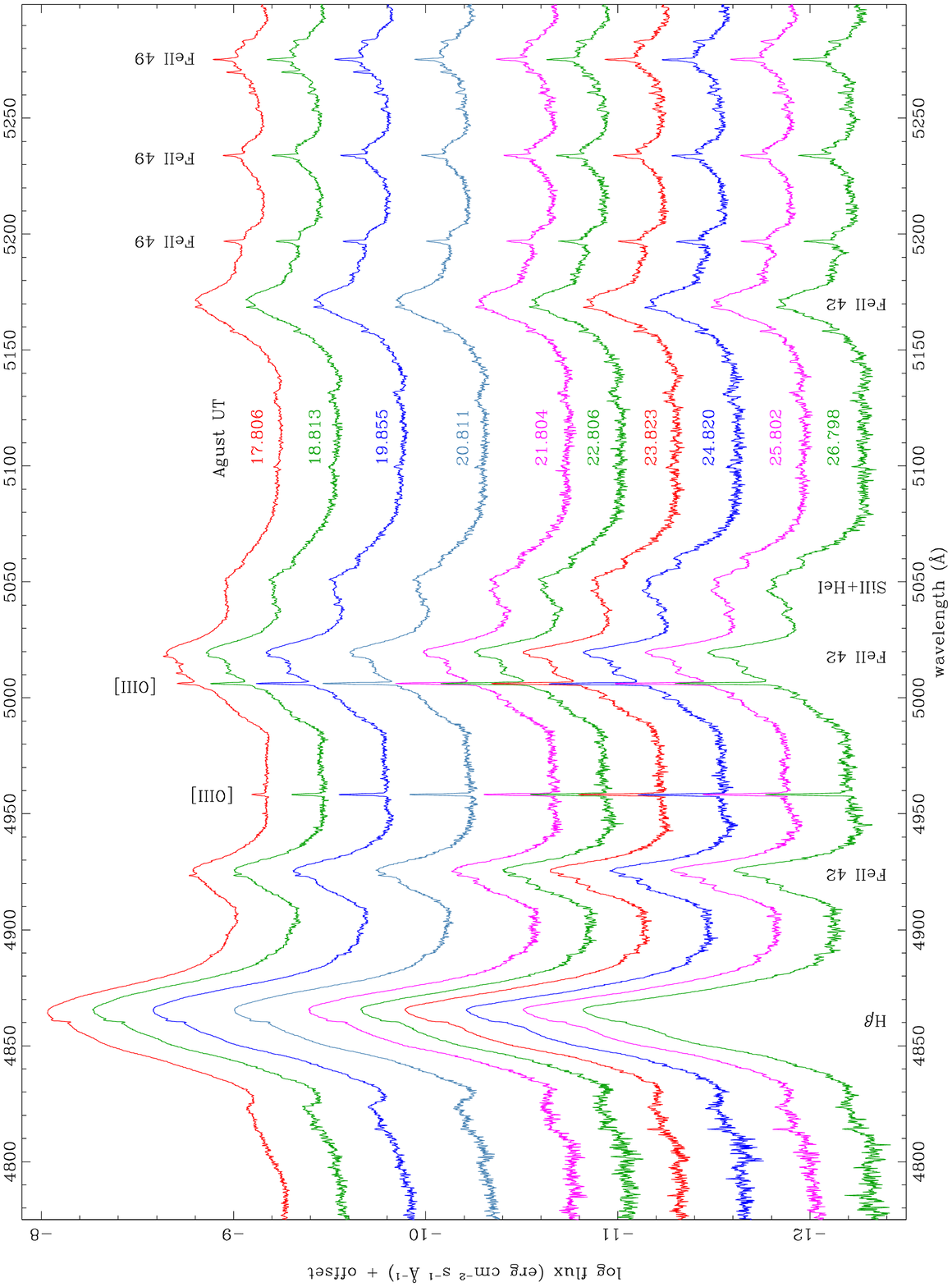}
\caption{Series of Varese 0.84m Echelle spectra covering the H$\beta$, FeII multiplet 42 and HeI 4941, 5015 
emission lines for the last ten days of our monitoring (the previous eight days are covered by preceeding Figure~6).
The logarithm of the flux is adopted as the ordinates and a shift is applied to avoid overplotting.}
\end{figure}

\clearpage
\begin{figure}
\includegraphics[angle=270,width=18cm]{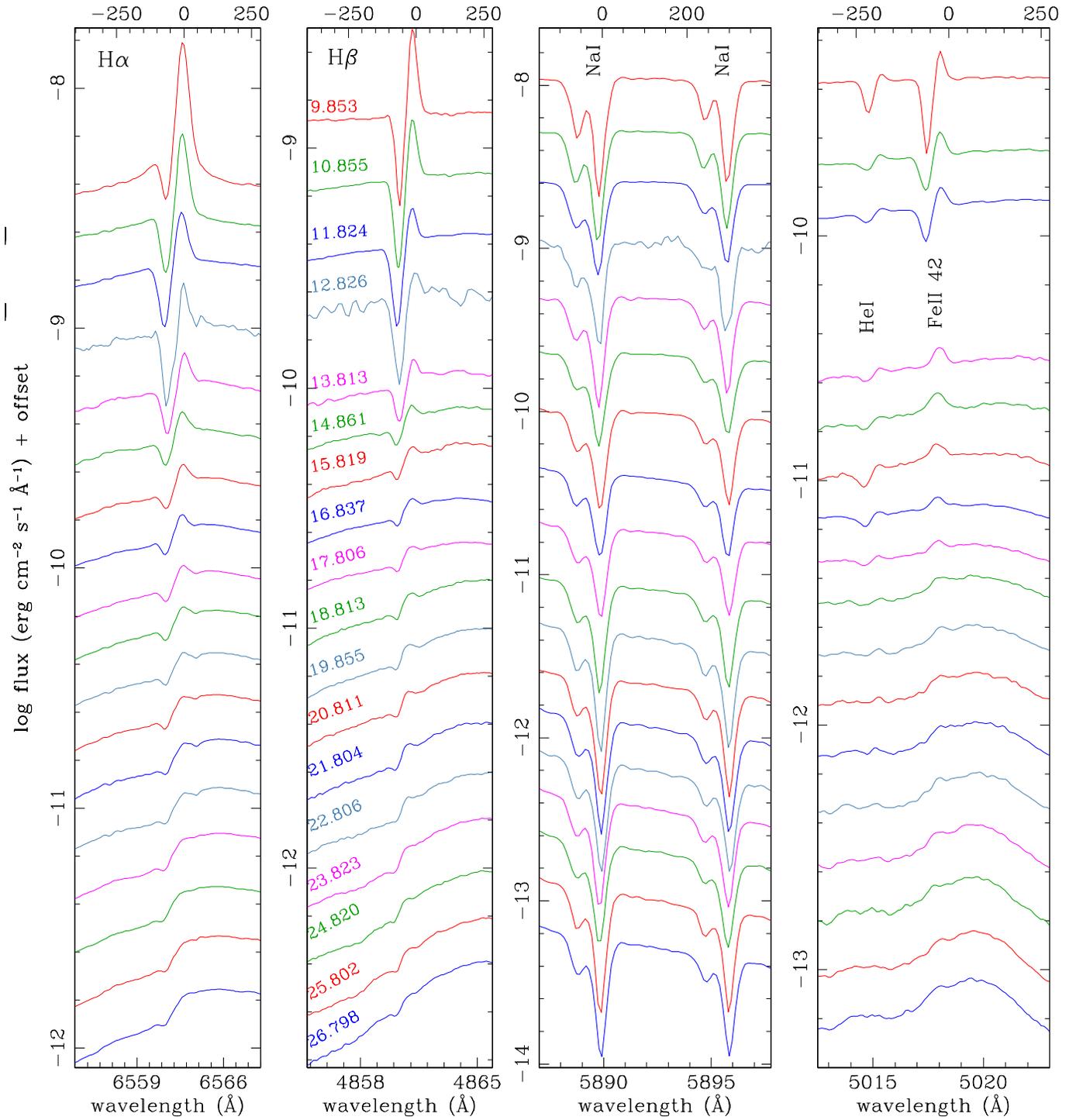}
\caption{Evolution of the narrow components for H$\alpha$, H$\beta$, HeI 5015 and FeII 5018, as well as
the NaI doublet from Varese 0.84m Echelle spectra. The abscissae at the top give the  heliocentric radial velocity (in
km/s).  The logarithm of the flux is adopted as the ordinates and a shift is applied to avoid overplotting.}
\end{figure}

\end{document}